\documentclass[conference]{IEEEtran}
\IEEEoverridecommandlockouts

\usepackage[T1]{fontenc}
\usepackage{cite}
\usepackage{amsmath,amssymb,amsfonts}
\usepackage{algorithm}
\usepackage{algpseudocode}
\usepackage{graphicx, multirow}
\usepackage{textcomp}
\usepackage[dvipsnames]{xcolor}
\usepackage[hyphens]{url}
\definecolor{myblue}{RGB}{72, 98, 168}
\usepackage[colorlinks=true, allcolors=myblue]{hyperref}
\usepackage{graphicx}
\usepackage{physics}
\usepackage{soul}
\usepackage{float}
\usepackage{siunitx}
\usepackage{ulem}
\usepackage{makecell}
\usepackage{bm}
\usepackage{tikz}
\usetikzlibrary{arrows.meta, positioning}
\def\BibTeX{{\rm B\kern-.05em{\sc i\kern-.025em b}\kern-.08em
    T\kern-.1667em\lower.7ex\hbox{E}\kern-.125emX}}

\newcommand{\name}[0]{CAbLECAR}

\newcommand{\Mod}[1]{\ \mathrm{mod}\ #1}
\DeclareUnicodeCharacter{2009}{\,}

\begin{document}





\title{CAbLECAR: efficiently scheduling QLDPC codes on a tileable spin qubit chip with shuttling
\thanks{This work is funded in part by the STAQ project under award NSF Phy-232580; in part by the US Department of Energy Office of Advanced Scientific Computing Research, Accelerated 
Research for Quantum Computing Program; and in part by the NSF Quantum Leap Challenge Institute for Hybrid Quantum Architectures and Networks (NSF Award 2016136), in part by the NSF National Virtual Quantum Laboratory program, in part based upon work supported by the U.S. Department of Energy, Office of Science, National Quantum 
Information Science Research Centers, and in part by the Army Research Office under Grant Number W911NF-23-1-0077. The views and conclusions contained in this document are those of the authors and should not be interpreted as representing the official policies, either expressed or implied, of the U.S. Government. The U.S. Government is authorized to reproduce and distribute reprints for Government purposes notwithstanding any copyright notation herein. FTC is the Chief Scientist for Quantum Software at Infleqtion.}
}

\author{\IEEEauthorblockN{Jason D. Chadwick}
\IEEEauthorblockA{\textit{Department of Computer Science} \\
\textit{University of Chicago}\\
Chicago, IL, USA \\
jchadwick@uchicago.edu}
\and

\IEEEauthorblockN{Frederic T. Chong}
\IEEEauthorblockA{\textit{Department of Computer Science} \\
\textit{University of Chicago}\\
Chicago, IL, USA \\
chong@cs.uchicago.edu}}

\maketitle

\thispagestyle{plain}
\pagestyle{plain}

\begin{abstract}
Semiconductor spin qubits are a promising platform for large-scale quantum computing, but have yet to take full advantage of the broad class of quantum low-density parity check (QLDPC) codes, which promise high encoding rates and efficient logic but require nonlocal connectivity between physical qubits. In this work, we investigate the implementation of QLDPC codes on a tileable, shuttling-based spin qubit architecture. By tailoring syndrome extraction circuits to the shuttling noise model, we significantly improve on previous surface code proposals and extend the feasible shuttling range of the architecture by 5-10$\times$, enabling the implementation of more complex codes with long-range interactions. Taking inspiration from the field of robotics, we develop a coordinated shuttle scheduling algorithm that supports arbitrary codes and use it to benchmark the logical performance of a variety of promising code families. We find that the optimized schedules are up to 86\% faster than hand-optimized schedules for certain code families. Through detailed circuit-level simulations, we identify specific QLDPC codes that improve upon prior surface code implementations by orders of magnitude, increasing encoding efficiency and reducing logical error rates. This work demonstrates the potential of shuttling-based spin qubit hardware platforms for scalable and efficient fault-tolerant quantum computation.
\end{abstract}


\section{Introduction}

\begin{figure}
    \centering
    \includegraphics[width=\linewidth] {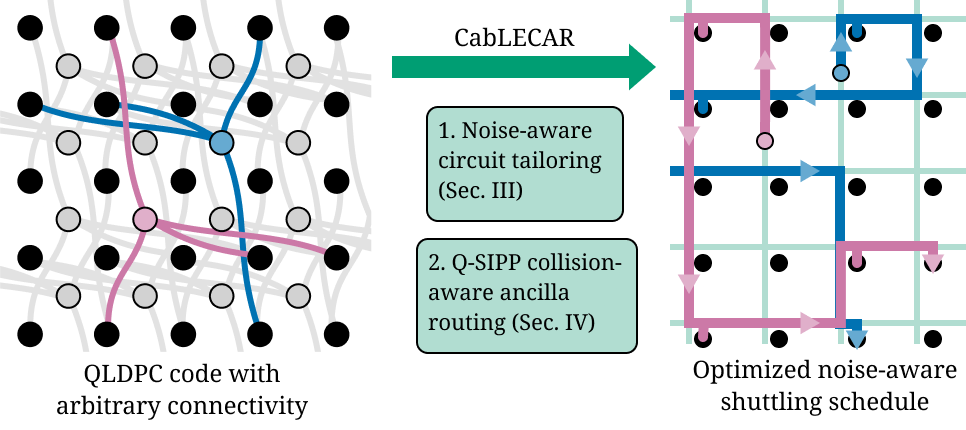}
    \caption{Overview of the \name{} framework. (Left) QLDPC codes demand complex, non-local parity checks, represented by arbitrary connections between data qubits (black) and ancilla qubits (colored). (Middle) To map these abstract requirements onto physical hardware, \name{} applies noise-aware circuit tailoring to mitigate shuttling-induced dephasing and utilizes a novel optimization algorithm Q-SIPP to compute collision-free routing paths. (Right) The output is a hardware-compatible, time-optimized shuttling schedule where mobile ancilla qubits physically route along a tileable grid architecture to interact with stationary data qubits. Applied to efficient QLDPC codes, these optimized schedules yield orders of magnitude improvement in QEC performance over the surface code.}
    \label{fig:hero}
\end{figure}

The realization of large-scale fault-tolerant quantum computing (FTQC) requires robust quantum error correction (QEC) to protect fragile quantum information from decoherence and operational errors. The surface code has long been the leading candidate for FTQC due to its high threshold, strictly nearest-neighbor connectivity, and simple logical operations \cite{kitaev_quantum_1997, dennis_topological_2002, fowler_surface_2012, horsman_surface_2012}. However, the surface code comes with a high physical qubit overhead to encode a single logical qubit, presenting a severe bottleneck for hardware scalability.

A promising alternative lies in the broader family of quantum low-density parity-check (QLDPC) codes. Many promising new QLDPC code families have been discovered that can achieve much higher encoding rates and lower logical error rates compared to the surface code \cite{bravyi_highthreshold_2024, scruby_highthreshold_2024, steffan_tile_2025, nakai_subsystem_2026}. Recent breakthroughs in methods for performing universal computation in QLDPC codes have further increased their abilities, allowing them to potentially be used not just for memory but for an entire quantum processor \cite{cross_linearsize_2024, xu_fast_2024, cowtan_parallel_2025, swaroop_universal_2025, sayginel_faulttolerant_2025, zhang_constantoverhead_2025, malcolm_computing_2025, xu_batched_2025, pecorari_addressable_2025, yang_spacetime-efficient_2026}. The primary barrier to implementing QLDPC codes on solid-state hardware is their requirement for complex, non-local connectivity between physical qubits---a feature that is native to modular atomic platforms but traditionally inaccessible to fixed two-dimensional solid-state arrays.

Semiconductor spin qubits offer a highly scalable pathway toward building millions of physical qubits, driven by their microscopic footprint and natural synergy with existing CMOS manufacturing processes \cite{maurand_cmos_2016, veldhorst_silicon_2017, li_crossbar_2018, xue_cmosbased_2021, neyens_probing_2024, steinacker_300_2024}. Although recent hardware milestones have demonstrated steady improvements in both array size \cite{borsoi_shared_2024, george_12spinqubit_2025, ha_twodimensional_2025, fernandez_de_fuentes_running_2026} and operational fidelity \cite{takeda_resonantly_2020, mills_twoqubit_2022, weinstein_universal_2023, wu_simultaneous_2025, broz_demonstration_2025, steinacker_industrycompatible_2025, madzik_operating_2025}, the realization of a dense monolithic grid of qubits remains an enormous engineering challenge. The strict requirements for control electronics, combined with severe wiring bottlenecks and signal crosstalk, force restrictive tradeoffs between array dimensions and control parallelism \cite{li_crossbar_2018, chadwick_manufacturable_2026}.

To bypass the limitations of monolithic density, modern architectural proposals favor sparse, modular designs connected by coherent electron shuttling \cite{boter_spiderweb_2022, kunne_spinbus_2024, otxoa_spinhex_2025}. By physically transporting qubits across the chip to perform entangling operations, these architectures free up critical on-chip real estate for essential routing and wiring. Although prior studies have successfully mapped the standard surface code onto these shuttling-based platforms \cite{yenilen_performance_2025}, demonstrating a viable path to near-term fault tolerance, such approaches are ultimately bottlenecked by the steep physical qubit overhead intrinsic to the surface code itself.

In this work, we demonstrate that the coherent shuttling capabilities of modern spin qubit architectures can be harnessed to unlock the benefits of QLDPC codes far exceeding the performance of the surface code. We present \name{} (\ul{C}oordinated \ul{A}ncillae for \ul{L}DPC codes with \ul{E}fficient \ul{C}ollision-\ul{A}ware \ul{R}outing), a comprehensive framework for compiling and optimizing arbitrary QLDPC codes on a tiled, shuttling-based spin qubit processor (Figure~\ref{fig:hero}). Recognizing that the shuttling range is physically limited by dephasing noise, we first introduce a noise-aware circuit tailoring technique that effectively cancels out shuttling-induced phase errors, dramatically extending the operational range of mobile ancilla qubits. To manage the complex web of non-local interactions in QLDPC codes, we adapt Safe Interval Path Planning (SIPP) from the field of autonomous robotics \cite{phillips_sipp_2011}, yielding Q-SIPP: a scalable, collision-aware scheduling algorithm capable of routing ancilla qubits for arbitrary stabilizer check structures.

Specifically, the core contributions of this paper are as follows:
\begin{itemize}
\item We show how simple circuit-level tailoring can mitigate the dominant dephasing errors associated with coherent shuttling, improving the resilience of the baseline surface code to shuttling errors by 5-10$\times$ and enabling longer-distance shuttles for more efficient QEC.
\item We develop Q-SIPP, a highly efficient routing algorithm that successfully coordinates collision-free paths for mobile ancilla qubits, achieving up to 85\% reduction in shuttling overhead compared to hand-optimized schedules.
\item We comprehensively benchmark the logical performance of several leading QLDPC code families, including tile, hypergraph-product-of-simplex, bivariate bicycle, and radial codes, against the baseline surface code on a shuttling-equipped architecture.
\item Through detailed circuit-level noise simulations, we identify the specific hardware noise regimes where QLDPC codes begin to yield orders-of-magnitude improvements in logical error rate and physical qubit encoding efficiency.
\end{itemize}

\section{Background}

\subsection{Quantum error correction}

\begin{figure}
    \centering
    \includegraphics[width=0.8\linewidth]{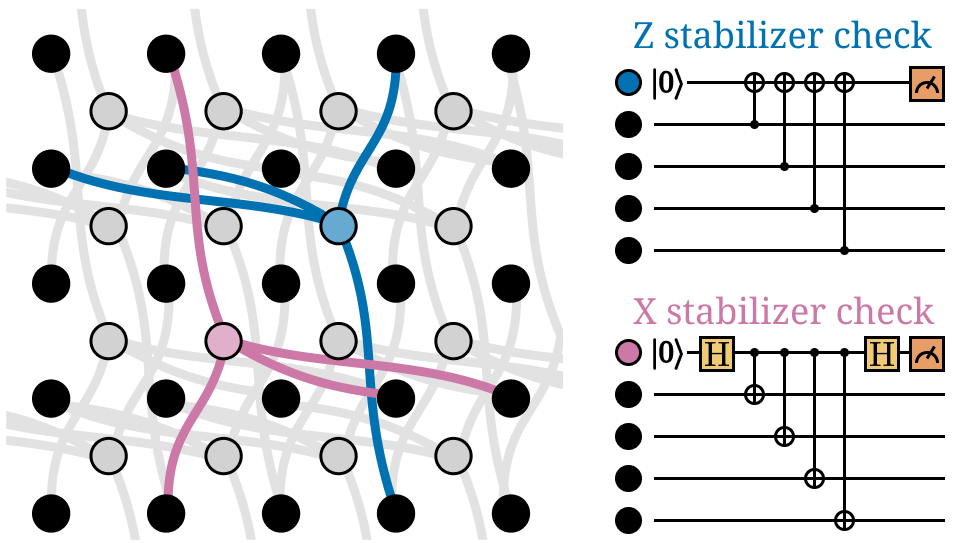}
    \caption{Depiction of the structure of a quantum error correction code. A code block consists of a group of data qubits (black) and the logical code state is stabilized by repeated measurement of X and Z stabilizers. Typically, each stabilizer is measured using an ancilla qubit, which is initialized, entangled with the relevant data qubits, and then measured. Example X and Z checks are shown in pink and blue. The measurement results must then be decoded to determine whether any correction is needed to maintain an error-free code state. In this figure, X and Z checks both have weight 4, meaning that each ancilla qubit interacts with four data qubits in a syndrome extraction (SE) cycle.}
    \label{fig:QEC-background}
\end{figure}

We briefly review the necessary fundamentals and terminology related to quantum error correction (QEC). For a more thorough background, we refer the reader to Refs. \cite{kitaev_quantum_1997, dennis_topological_2002, fowler_surface_2012}.

Quantum error correction is the process of encoding a small number of long-lived logical qubits in a large number of noisy physical qubits. In a standard QEC framework, a code block consists of two distinct types of qubit: data qubits, which collectively store the logical quantum state(s) and ancilla qubits, which are used to extract error information without collapsing the logical state. The logical code state is protected by enforcing a set of parity checks, which requires the repeated measurement of X and Z stabilizers. Figure~\ref{fig:QEC-background} shows an abstract QEC code, with the data qubits (black) arranged in a grid and the ancilla qubits (gray and colored) amongst them. Each ancilla qubit is associated with an X or Z check, which is the measurement of the joint X or Z observable across a subset of the data qubits. Example X and Z checks are shown in pink and blue, along with the quantum circuits corresponding to one measurement of the stabilizers. An important parameter of these checks is their \textit{weight}, which is the number of data qubits involved in the check. The weight of a code refers to the maximum weight of any individual parity check in the code.

The process of measuring these stabilizers is known as syndrome extraction (SE). During a typical syndrome extraction cycle, each ancilla qubit is initialized, entangled with the relevant data qubits, and then measured, as shown in the circuits in Figure~\ref{fig:QEC-background}. The measurement results must then be decoded with a classical decoding algorithm, typically in batches of $d$ SE rounds, to determine whether any correction is needed to maintain an error-free code state. 

The performance scaling of a QEC code can be quantified with the code distance $d$, which is the minimum number of physical faults that can cause a logical error. If all physical errors occur with probability $p$, the expected logical error rate scales as $p_L \propto (p/p^*)^{(d+1)/2}$, where the parameter $p^*$ is known as the code threshold and determines the maximum value of $p$ such that increasing the code distance can arbitrarily suppress errors. QEC codes are often identified by their parameters $[[n,k,d]]$, where $n$ is the number of data qubits in a code block and $k$ is the number of logical qubits in the code block. For example, the surface code family consists of $[[d^2, 1, d]]$ codes. The quantity $n/k$ is known as the encoding rate.

\subsection{Semiconductor spin qubits}

Semiconductor spin qubits have emerged as one of the most promising solid-state hardware platforms for large-scale quantum computation. These qubits typically encode quantum information in the spin states of individual electrons confined within semiconductor quantum dots. They offer compelling physical advantages for scalability, including a \SI{}{\micro\meter}-scale footprint, the potential for operation at elevated temperatures, and direct compatibility with advanced CMOS manufacturing processes. Recent state-of-the-art demonstrations have shown the operation of two-dimensional arrays with capacities up to 16 qubits \cite{borsoi_shared_2024, ha_twodimensional_2025, fernandez_de_fuentes_running_2026}, and arrays with the capability to hold over 100 qubits have been successfully fabricated \cite{henry_2d_2025}. Gate and readout fidelities nearing or exceeding 99.9\% have been demonstrated in a variety of hardware experiments \cite{takeda_resonantly_2020, mills_twoqubit_2022, weinstein_universal_2023, wu_simultaneous_2025, broz_demonstration_2025, steinacker_industrycompatible_2025, madzik_operating_2025}. $T_1$ relaxation times for spin qubits are generally extremely high, on the order of seconds \cite{tyryshkin_electron_2012}. $T_2^*$ dephasing times are lower, on the order of hundreds of \SI{}{\micro\second}, but can be extended to $T_2$ times of tens of \SI{}{\milli\second} with dynamical decoupling techniques \cite{veldhorst_addressable_2014}. For a thorough review of semiconductor spin qubits, we refer the interested reader to Ref. \cite{burkard_semiconductor_2023}.

Despite these physical advantages, the operation of dense, monolithic two-dimensional arrays of spin qubits is severely complicated by wiring bottlenecks, signal crosstalk, and the physical space required for control electronics. These challenges often force a difficult tradeoff between the width of the array and the control parallelism \cite{li_crossbar_2018, chadwick_manufacturable_2026}. To circumvent these dense connectivity challenges, recent architectural proposals have shifted toward modular, sparse arrays equipped with coherent qubit shuttling or long-range couplers \cite{boter_spiderweb_2022, kunne_spinbus_2024, otxoa_spinhex_2025}. In these designs, physical qubits are physically transported across the chip to perform entangling operations, leaving large areas of the chip available to allow for scalable wiring. Recently, a proof-of-principle weight-four parity check with a mobile ancilla qubit has been successfully demonstrated in silicon \cite{undseth_weight-four_2026}.

\begin{figure}
    \centering
    \includegraphics[width=\linewidth]{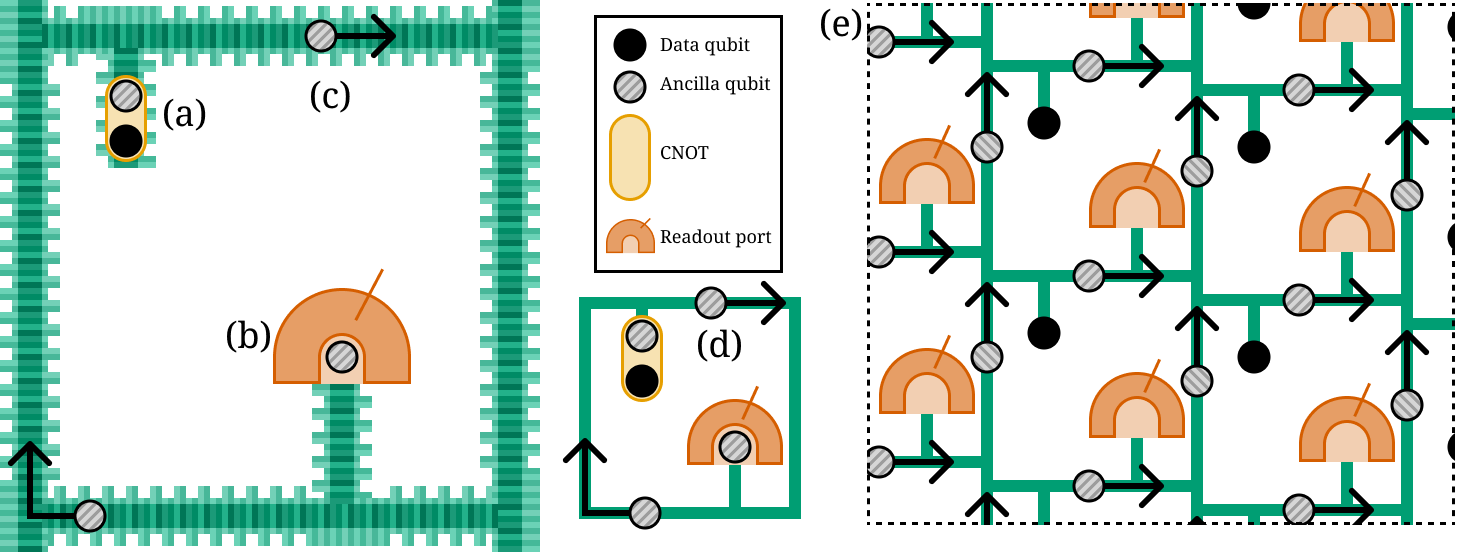}
    \caption{The tiled unit cell architecture considered in this work. On the left is a closeup of one unit cell, which has (a) one interaction zone where gates are performed, (b) one readout port, where a qubit can be measured, and (c) shuttling channels, where qubits can travel within and between unit cells. (d) shows a simplified version of the unit cell, and (e) shows how the unit cells can be tiled into a grid-like arrangement.}
    \label{fig:unit-cell}
\end{figure}

The hardware target we consider is heavily inspired by the SpinBus proposal by Kunne et al. \cite{kunne_spinbus_2024}. The chip is built from unit cells as shown in Figure~\ref{fig:unit-cell}, where each unit cell contains one interaction zone (where two qubits can be entangled) and one readout zone (where a qubit can be initialized or measured). The unit cell is tiled to make a large array, with shuttling channels connecting the unit cells to each other. The crucial capability enabling this proposal, long-distance coherent shuttling, has been demonstrated in several test chips \cite{seidler_conveyormode_2022, zwerver_shuttling_2023, xue_sisige_2024}. In particular, shuttling an effective distance of \SI{10}{\micro\meter}, the expected edge length of one unit cell, was achieved with an overall fidelity of 99.5\% and a speed of over \SI{50}{\meter/\second} \cite{smet_highfidelity_2024}. Theoretical analysis predicts that these shuttling fidelities could reach as low as $10^{-5}$ (and speeds as high as hundreds of \SI{}{\meter/\second}) through pulse shaping \cite{oda_suppressing_2026}. An important feature of coherent electron shuttling that we leverage in this work is that shuttling errors are heavily biased towards dephasing, with bit-flip errors occurring several orders of magnitude less frequently \cite{foster_dephasing_2025, smet_highfidelity_2024}.

\section{Extending shuttling range with noise-aware circuit tailoring}\label{sec:tailoring}

\begin{figure}
    \centering
    \includegraphics[width=\linewidth]{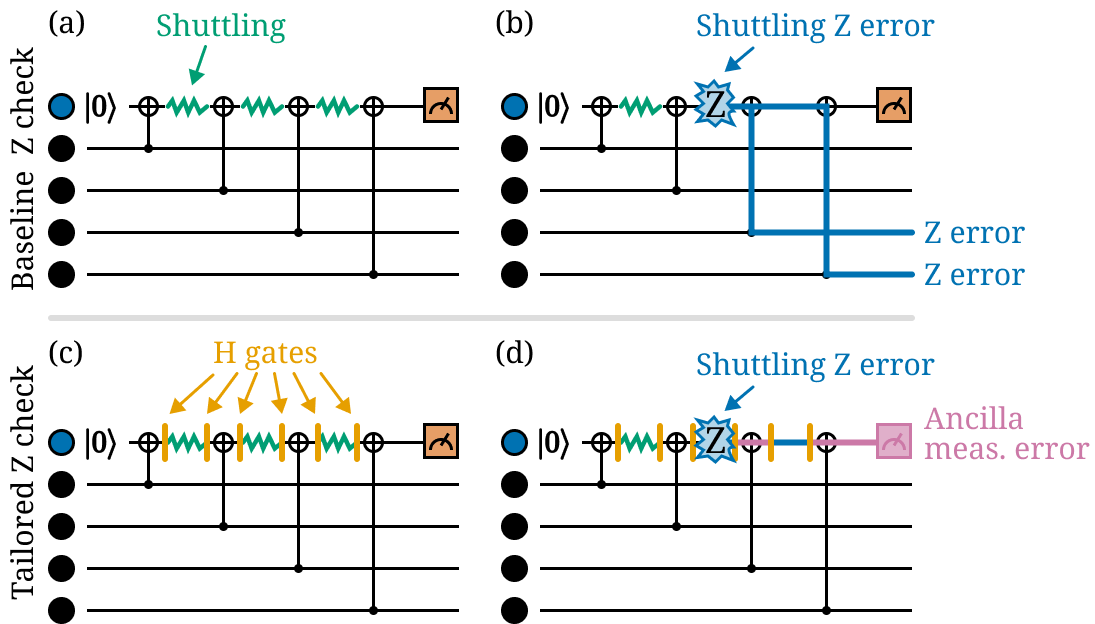}
    \caption{Tailoring syndrome extraction circuits to mitigate shuttling noise. (a) Ancilla qubits are shuttled between CNOTs with data qubits. Shown here is a Z ancilla qubit, which is sensitive to shuttling errors. (b) Shuttling errors are heavily biased towards Z errors (dephasing). When such an error occurs, it propagates through the following CNOTs and manifests as one or more Z errors on the data qubits. (c) The tailored version of the circuit inserts H gates before and after each shuttling period. (d) This switches the Z error to an X error during each of the remaining CNOTs, preventing error propagation. The consequence is an X error on the ancilla qubit at the end of the circuit, which causes an incorrect measurement result.}
    \label{fig:tailoring}
\end{figure}

We first describe a simple optimization to standard syndrome extraction circuits that greatly increases resilience to shuttling errors, enabling longer-distance stabilizer checks. We exploit the strong noise bias of the shuttling error channel, which is almost entirely dephasing errors \cite{foster_dephasing_2025, smet_highfidelity_2024}. Inspecting the Z stabilizer syndrome extraction circuit in Figure~\ref{fig:tailoring}a-b, we see that a shuttling error on a Z ancilla qubit propagates to the data qubits that it interacts with afterwards, leading to one or more Z errors. Given that only the ancilla qubits are shuttled, we can mitigate this error propagation by simply appending Hadamard gates before and after every shuttling period. In the noiseless case, each pair of Hadamards is equivalent to the identity, so the syndrome extraction circuit still logically functions identically. As shown in Figure~\ref{fig:tailoring}d, if a shuttling error occurs during a shuttling period, each Hadamard gate will toggle the error between a Z error and an X error. This prevents error propagation to the data qubits, instead causing an X error on the ancilla qubit, which becomes a bit-flip measurement error when the ancilla is measured. The X ancilla checks do not need to be modified because dephasing errors on X ancillae do not propagate to the data qubits.

\begin{figure}
    \centering
    \includegraphics[width=\linewidth]{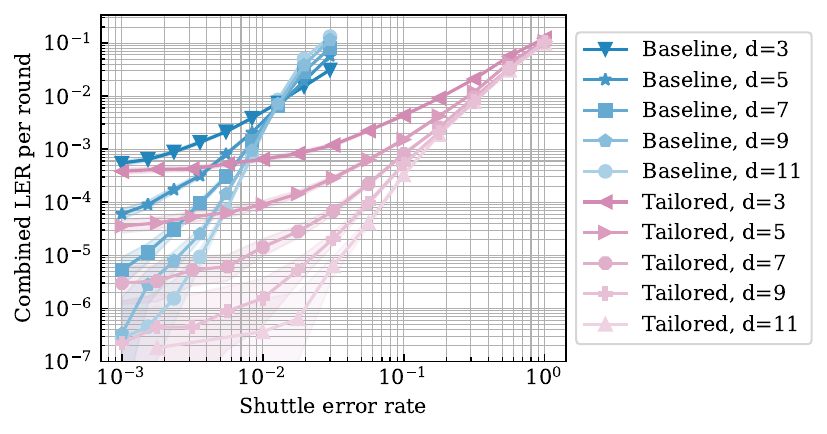}
    \caption{Performance of the surface code shuttling schedule on the unit cell architecture \cite{yenilen_performance_2025} under varying shuttling noise, with and without \name{} circuit tailoring. Tailoring allows the surface code to tolerate 5-10$\times$ higher shuttle error rates on the ancilla qubits. This increased noise resilience paves the way for considering more complex codes with longer shuttling paths.}
    \label{fig:tailoring-results}
\end{figure}

This optimization can be immediately applied to the prior surface code schedule \cite{yenilen_performance_2025} to improve its resilience to shuttling errors. Figure~\ref{fig:tailoring-results} shows the combined X and Z memory error rate of surface codes of various distance as a function of the shuttling error rate with and without this circuit tailoring. Gate and measurement error rates are held constant at $10^{-3}$. The bus error rate is the chance of a dephasing error per unit edge (in the surface code circuit, each ancilla qubit travels a total of four edge lengths per SE round). In the baseline circuit (blue) we observe a clear threshold around a bus error rate of $10^{-2}$, and see that shuttling error rates closer to $10^{-3}$ are required to yield strong suppression of logical errors with increasing code distance. On the other hand, the tailored circuit (pink) can achieve similar logical error rates at a 5-10$\times$ higher bus error rate.

Because this circuit tailoring converts shuttling errors to ancilla measurement errors, increasing the temporal code distance by decoding over a window with more than the standard $d$ SE rounds can further improve the per-round logical error rate. While this may allow the use of spatially-smaller codes to achieve the same logical performance, increasing the temporal distance of the code lengthens the logical clock cycle time, which may lead to a relative slowdown. These spacetime cost tradeoffs deserve a more in-depth exploration that is beyond the scope of this work; here we choose to simply decode over $d$ SE rounds as usual and leave a more thorough study for future work.

We have shown that the shuttling noise circuit tailoring technique significantly improves the performance of the prior surface code schedule. Crucially, by relaxing the tolerable shuttling error threshold by nearly an order of magnitude, this tailoring technique physically permits the extended routing distances required by QLDPC codes. This widened error margin serves as the essential hardware foundation that makes the complex, long-range ancilla routing in \name{} feasible.

\section{Quantum Safe Interval Path Planning (Q-SIPP) for Syndrome Extraction}

\begin{figure}
    \centering
    \includegraphics[width=\linewidth]{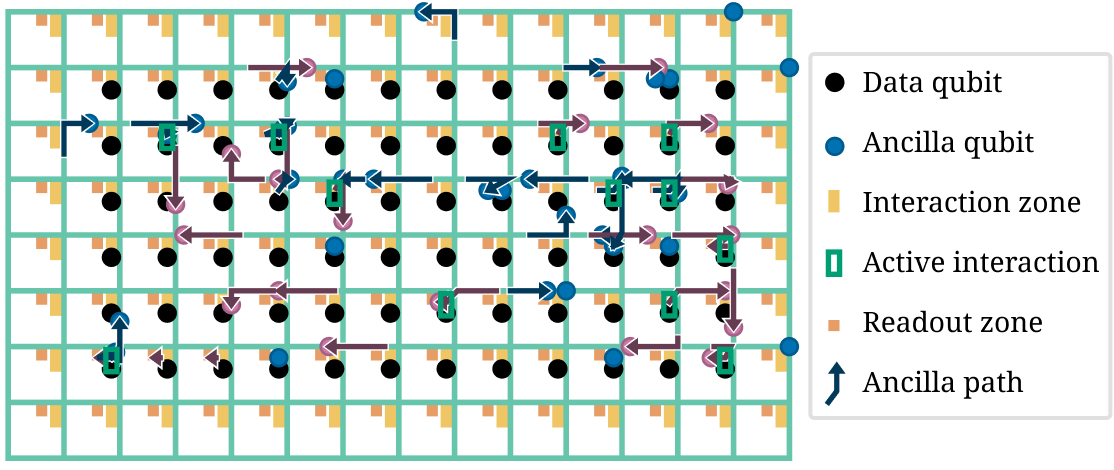}
    \caption{Snapshot of optimized ancilla movement schedule for the [[72, 12, 6]] bicycle code. Data qubits (black) are in fixed positions, one per unit cell. Ancilla qubits (blue and pink) are shuttled between data qubits, performing CNOTs along the way. Colored arrows show ancilla movement over a 1\si{\micro\second} time period.}
    \label{fig:schedule-example}
\end{figure}

Most prior proposals for the implementation of error correction codes with mobile qubits have resorted to hand-optimized movement patterns, exploiting the structure of particular code families that allow qubit movements to be synchronized \cite{cai_looped_2023, viszlai_matching_2023, xu_constantoverhead_2024, siegel_early_2024, liu_coniq_2025}. However, this approach requires significant effort to apply to a new code family and is impractical for codes that have a less convenient structure. In this section, we describe a novel optimization approach that enables the implementation of arbitrary high-rate codes on shuttling-equipped architectures.

Scheduling the shuttling movements on our device takes the form of a simple problem: each ancilla qubit has a set of data qubits $\{d_i\}$ at fixed locations that it must visit. This list may either be strictly ordered to carry out a particular hook-error-avoiding syndrome extraction circuit, such as for the surface code, or may be unordered, allowing more freedom to choose the optimal ancilla order on the fly. For example, hypergraph product codes are known to allow any ordering of CNOTs in an SE circuit without a reduction in the effective code distance \cite{manes_distance-preserving_2025}. Upon arriving at a data qubit, the two qubits are entangled, and then the ancilla qubit either moves on to its next task or to a readout port if it has completed its tasks. The objective is to schedule shuttling operations to minimize the overall duration of the syndrome extraction round. Figure~\ref{fig:schedule-example} shows a snapshot of a \name{}-optimized schedule for the [[72,12,6]] bicycle code \cite{bravyi_highthreshold_2024}, where we show the paths of ancilla qubits (pink and blue) over a \SI{1}{\micro\second} time interval.

The ancilla qubits are scheduled one-by-one in a predefined order, where each previously-scheduled ancilla qubit is treated as a dynamic obstacle that must be avoided. This leads to a problem space where each device component (shuttle channel, interaction zone, or readout port) has time-dependent availability. Standard path planning in dynamic environments often suffers from an explosion in the search space because it must add time as a fully discretized, independent dimension. To avoid this, \name{} employs a quantum-tailored variant of Safe Interval Path Planning (SIPP) \cite{phillips_sipp_2011} that we refer to as Quantum-SIPP (Q-SIPP).

SIPP is based on the observation that the number of distinct time intervals where each component is available (unoccupied) is finite and generally very small. Therefore, instead of fully discretizing time, we can simply keep track of \textit{safe intervals} for each component. A safe interval is defined as a time period where a component is unoccupied, bounded by occupations on either side, unless it is the final interval which may extend to infinity. By representing states with safe intervals rather than discrete timesteps, the search space is significantly compressed, allowing for highly efficient planning \cite{phillips_sipp_2011}. Each Q-SIPP schedule optimization performed in the preparation of this manuscript, for codes with up to 512 data qubits and 494 ancilla qubits, ran in under 5 minutes on a single CPU core.

\subsection{State Space Formulation}
To accommodate the specific rules of the quantum hardware, the standard SIPP configuration space is expanded. A state $S$ in Q-SIPP is defined as the tuple $\langle loc, comp, i, T \rangle$:
\begin{itemize}
    \item $loc$: The spatial coordinates of the qubit on the chip.
    \item $comp$: The currently-occupied hardware component type (e.g. shuttle channel, interaction zone, readout).
    \item $i$: The index of the safe interval for this specific location and component.
    \item $T$: A boolean tuple tracking task progress (specifically, which data qubits the ancilla has successfully interacted with).
\end{itemize}

By tracking time with safe intervals, we do not need to record the precise time that the qubit arrives at or leaves a location, significantly reducing problem complexity. A compiled shuttling schedule with exact timings can be extracted from the final solution path.

\begin{algorithm}[t]
\caption{Q-SIPP Search}
\label{alg:qsipp}
\begin{algorithmic}[1]
\Require Start state $s_{start}$, goal conditions
\Ensure Optimal time-ordered schedule
\State $g(s_{start}) \gets 0$; $OPEN \gets \{s_{start}\}$
\While{$OPEN \neq \emptyset$}
    \State $s \gets \arg\min_{s' \in OPEN} f(s')$
    \State $OPEN \gets OPEN \setminus \{s\}$
    \If{\Call{IsComplete}{$s.T$} \textbf{and} $s.comp = \text{READOUT}$}
        \State \Return \Call{TracePath}{$s$}
    \EndIf
    \For{\textbf{each} $s' \in \Call{GetSuccessors}{s}$}
        \State $t_{arr} \gets \max(g(s) + c(s,s'), s'.i_{start})$
        \If{$t_{arr} < g(s')$}
            \State $g(s') \gets t_{arr}$
            \State $f(s') \gets g(s') + \Call{Heuristic}{s'}$
            \State insert or update $s'$ in $OPEN$ with $f(s')$
        \EndIf
    \EndFor
\EndWhile
\State \Return failure
\end{algorithmic}
\end{algorithm}

\subsection{A* Search Loop}
Algorithm \ref{alg:qsipp} outlines the core A* search operating over these states. It evaluates nodes based on the objective function $f(s) = g(s) + h(s)$, where $g(s)$ tracks the cost of the shortest known path from the start state to $s$ and the heuristic function $h(s)$ estimates the remaining cost to task completion. At each step, the state $s$ with the minimum objective function is removed from the $OPEN$ set, which is implemented as a priority queue. The search successfully terminates when it expands a goal state, which is a state where all tasks in $T$ are complete and the qubit is safely resting in a READOUT component. SIPP guarantees that this schedule will be time-optimal, given fixed preexisting safe interval constraints. If the chosen state is not a goal state, Q-SIPP finds all valid successor states of $s$ using the \textsc{GetSuccessors} function described in the following subsection. These states are added to the $OPEN$ set.

\begin{algorithm}[t]
\caption{\textsc{GetSuccessors}($s$)}
\label{alg:successors}
\begin{algorithmic}[1]
\Require Current state $s = \langle loc, comp, i, T \rangle$
\State $succ \gets \emptyset$
\State $P \gets \text{PendingDataLocations}(s.T)$
\If{$s.comp = \text{INTERSECTION}$}
    \State $succ \gets succ \cup \Call{GetValidShuttles}{s}$
\EndIf
\For{\textbf{each} $c \in \text{PossibleComponents}\ \setminus \{s.comp\}$}
    \State $succ \gets succ \cup \Call{GetValidDisplacements}{s, c}$
\EndFor
\If{$s.comp = \text{INTERACTION}$ \textbf{and} $s.loc \in P$}
    \If{$s.i_{end} \geq g(s) + t_{CX}$}
        \State $T' \gets \Call{UpdateTask}{s.T, s.loc}$
        \State $s' \gets \langle s.loc, s.comp, s.i, T' \rangle$
        \State insert $s'$ into $succ$ with cost $f(s') = f(s) + t_{CX}$
    \EndIf
\EndIf
\State \Return $succ$
\end{algorithmic}
\end{algorithm}

\subsection{Successor Generation and Transitions}
Algorithm \ref{alg:successors} determines how the successor states are chosen upon expansion of a state. There are several actions an ancilla qubit can take to transition to a new state:
\begin{enumerate}
    \item \textbf{Shuttle:} If the current state's component is an \textsc{Intersection}, the qubit can shuttle to an adjacent \textsc{Intersection}. This transition is valid if the intermediate \textsc{Channel} and destination \textsc{Intersection} have compatible safe intervals, meaning that the ancilla qubit can leave the current \textsc{Intersection}, shuttle through the \textsc{Channel}, and arrive at the new \textsc{Intersection} during these safe intervals. There may be multiple distinct transition states if the destination \textsc{Intersection} has multiple reachable safe intervals.
    \item \textbf{Displace:} Transition between hardware component layers at its current location (e.g., moving from an \textsc{Intersection} into an \textsc{InteractionZone} or \textsc{ReadoutZone}).
    \item \textbf{Execute Gate:} Perform a CX gate with a data qubit, if resting in an \textsc{InteractionZone} corresponding to a pending data qubit, resulting in a new state with updated task progress tracker $T$.
\end{enumerate}

Idling is handled implicitly via the safe interval formalism and does not need to be directly encoded as a state transition.

\begin{figure}
    \centering
    \includegraphics[width=\linewidth]{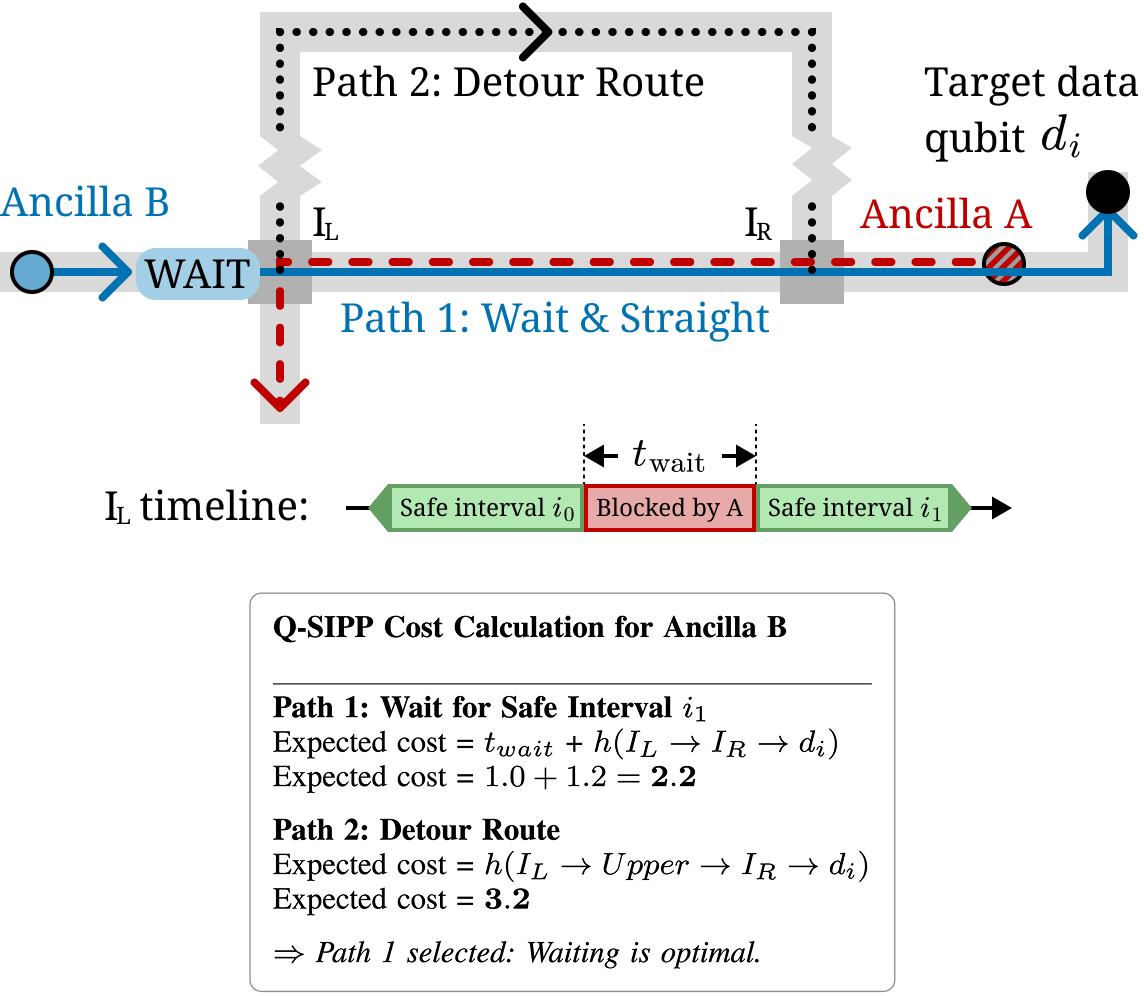}
    
    \caption{A visualization of the Q-SIPP scheduling process resolving a spatial collision. Ancilla B must route to $d_i$ but its shortest path through $I_L$ is temporarily blocked by Ancilla A. Instead of forcing B to take a collision-free detour, Q-SIPP projects the timeline of $I_L$ into continuous safe intervals. By calculating the heuristic cost $h(s)$ of the detour versus the known penalty $t_{\text{wait}}$ to utilize the next available safe interval ($i_1$), the algorithm correctly determines that waiting at $I_L$ yields the time-optimal schedule.}
    \label{fig:SIPP}
\end{figure}

\subsection{Heuristic function}
To ensure that the A* search remains highly directed, Algorithm \ref{alg:heuristic} provides an admissible heuristic that estimates the remaining cost-to-goal. For a given state, the heuristic function identifies all pending data qubits remaining in the task tuple $T$. It then utilizes a traveling salesman solver to determine the shortest possible path to complete the remaining tasks, ignoring dynamic obstacles entirely. The overall estimated cost is the total duration of shuttling along this path plus the durations of performing the required displacements and CX gates at each stop.

The resulting heuristic cost dynamically incorporates the minimum required shuttling time between the remaining data qubits, the hardware-specific displacement delays required to interact with them, and the strict durations of the remaining quantum gates. Because this heuristic always gives the shortest possible collision-free time to the goal, it is consistent and satisfies the triangle inequality, so the completeness claims from Ref.~\cite{phillips_sipp_2011} hold. The codes we evaluate in this work all have check weight $\leq$ 8, for which the exact TSP solver runs extremely quickly. For codes with a sufficiently-large stabilizer check weight, heuristic TSP solvers may be required instead of exact methods, which would remove the completeness guarantee, but Q-SIPP could still generate valid solution schedules. Although we have not exhaustively mapped its scalability bounds, preliminary results show that Q-SIPP successfully generates high-quality schedules for check weights up to 12 when paired with approximate TSP solvers.

\begin{algorithm}[t]
\caption{\textsc{Heuristic}($s$)}
\label{alg:heuristic}
\begin{algorithmic}[1]
\Require Current state $s$
\Ensure Estimated minimum remaining time
\State $P \gets \text{PendingDataLocations}(s.T)$
\If{$P = \emptyset$} \Comment{All tasks complete}
    \If{$s.comp = \text{READOUT}$}
        \State \Return $0$
    \Else
        \State \Return $t_{displace}$
    \EndIf
\EndIf
\State $cost \gets 0$
\State $cur \gets s.loc$
\If{$s.comp = \text{INTERACTION}$ \textbf{and} $cur \in P$}
    \State $cost \gets cost + t_{CX} + t_{displace}$
    \State $P \gets P \setminus \{cur\}$
\EndIf
\If{$s.comp \neq \text{INTERACTION}$ \textbf{and} $cur \notin P$}
    \State $cost \gets cost + t_{displace}$
\EndIf
\State $\pi \gets \Call{SolveTSP}{cur, P}$
\For{\textbf{each} segment $(u, v) \in \pi$}
    \State $cost \gets cost + \Call{Dist}{u,v} \cdot t_{shuttle} + t_{CX} + 2 \cdot t_{displace}$
\EndFor
\State \Return $cost$
\end{algorithmic}
\end{algorithm}


\subsection{Extensions}

The Q-SIPP algorithm is quite flexible and can easily be extended to consider alternative problem formulations. For example, we make a small modification to account for the circuit tailoring described in Section~\ref{sec:tailoring} by replacing each CX gate with a sequence of Hs and CXs (for Z ancilla qubits only). This can be incorporated by simply changing $t_{CX}$ to $t_{CX}+2t_{H}$ in Algorithms~\ref{alg:successors} and \ref{alg:heuristic}.

Another key feature of the Q-SIPP algorithm is its adaptability to different hardware layouts. For example, the original SpinBus proposal \cite{kunne_spinbus_2024} involves a hexagonal grid of unit cells instead of the square grid that we consider in this work. Alternatively, due to wiring and chip layout constraints, it may be beneficial to adopt a sparser layout with dense clusters of data qubits connected by long-range shuttling channels. In either case, Q-SIPP can be easily modified to optimize over arbitrary shuttling connectivities.

\section{Comparing code performance}\label{sec:eval}

We choose several promising families of QLDPC codes to compile with \name{} and analyze. The chosen code instances are summarized in Table~\ref{tab:codes}. In addition to the surface code, which we use as a baseline, we study the following code families: \textbf{tile codes}, which maintain surface-code-like locality but offer improved encoding efficiency \cite{steffan_tile_2025}; \textbf{hypergraph product codes of simplex codes}, which offer compelling automorphism-based logical operations \cite{malcolm_computing_2025, yang_spacetime-efficient_2026}; high-threshold and low check weight \textbf{bivariate bicycle codes} \cite{bravyi_highthreshold_2024}; and \textbf{radial codes}, which offer compelling performance and single-shot decodability \cite{scruby_highthreshold_2024}, but do not easily translate to hardware implementations due to their more complex check structures. The table lists each code with its key parameters $[[n,k,d]]$, where $n$ is the number of data qubits in a code block, $k$ is the number of encoded logical qubits, and $d$ is the code distance. The encoding rate $n/k$ is therefore the number of physical qubits required to encode each logical qubit.

\begin{table}[t]
\centering
\renewcommand{\arraystretch}{1.2}
\begin{tabular}{llrll}
Code type & $[[n, k, d]]$ & $n/k$ & Check weight &  Source \\ \hline
Surface & [[9, 1, 3]]   & 9.0 & 4 & \cite{bombin_optimal_2007} \\
Surface & [[25, 1, 5]]  & 25.0 & 4 & \cite{bombin_optimal_2007} \\
Surface & [[49, 1, 7]]  & 49.0 & 4 & \cite{bombin_optimal_2007} \\
Surface & [[81, 1, 9]]  & 81.0 & 4 & \cite{bombin_optimal_2007} \\
Surface & [[121, 1, 11]]& 121.0 & 4 & \cite{bombin_optimal_2007} \\
Tile    & [[288, 8, 12]] & 36.0 & 6 & \cite{steffan_tile_2025} \\
Tile    & [[288, 8, 14]] & 36.0 & 8 & \cite{steffan_tile_2025} \\
Tile    & [[288, 18, 13]]& 16.0 & 8 & \cite{steffan_tile_2025} \\
Tile    & [[512, 18, 19]]& 28.4 & 8 & \cite{steffan_tile_2025} \\
Simplex & [[98, 18, 4]]  & 5.4 & 6 & \cite{malcolm_computing_2025} \\ 
Simplex & [[450, 32, 8]] & 14.1 & 6 & \cite{malcolm_computing_2025} \\ 
BB      & [[72, 12, 6]]  & 6.0 & 6 & \cite{bravyi_highthreshold_2024} \\
BB      & [[144, 12, 12]]& 12.0 & 6 & \cite{bravyi_highthreshold_2024} \\
BB      & [[288, 12, 18]]& 24.0 & 6 & \cite{bravyi_highthreshold_2024} \\
Radial  & [[144, 8, 12]] & 18.0 & 6 & \cite{scruby_highthreshold_2024} \\
Radial  & [[198, 8, 14]] & 24.8 & 6 & \cite{scruby_highthreshold_2024} \\
Radial  & [[192, 18, 12]]& 10.7 & 8 & \cite{scruby_highthreshold_2024} \\
Radial  & [[256, 18, 16]]& 14.2 & 8 & \cite{scruby_highthreshold_2024} \\
Radial  & [[352, 18, 20]]& 19.6 & 8 & \cite{scruby_highthreshold_2024} \\
\end{tabular}

\hspace{0.5ex}

\caption{Code constructions studied in this work}
\label{tab:codes}
\end{table}

\begin{figure}
    \centering
    \includegraphics[width=\linewidth]{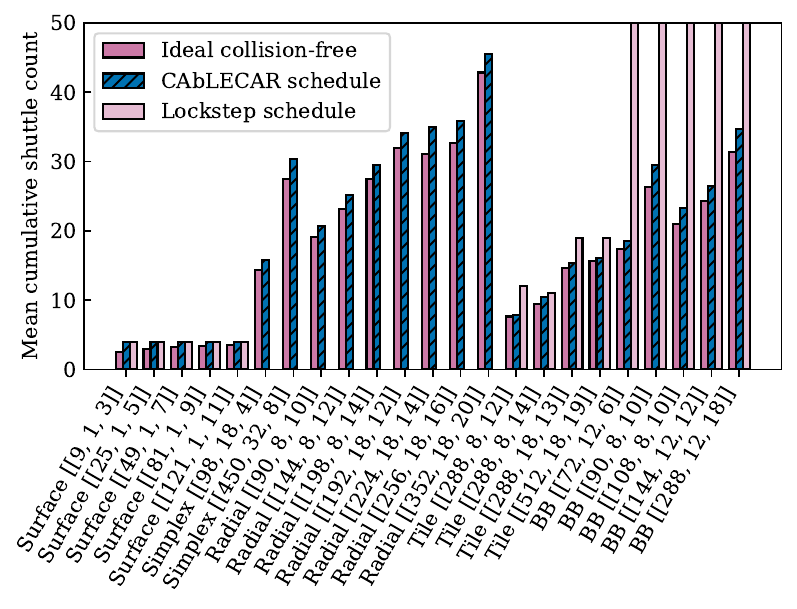}
    \caption{Comparing average shuttle distance for \name{} schedules compared to ideal minimum shuttle counts (assuming ancilla qubits can move freely past each other without collisions). We also compare to a ``lockstep'' schedule (all ancilla qubits moving together in a pattern) for codes where such a schedule exists. We find that \name{} outperforms all lockstep schedules and is close to the collision-free ideal, with a minor 12.5\% overhead (geomean). \name{} reduces shuttle cost for Tile and BB codes by 19\% and 86\%, respectively, compared to hand-optimized lockstep schedules.}
    \label{fig:shuttle-count}
\end{figure}

\begin{figure}
    \centering
    \includegraphics[width=\linewidth]{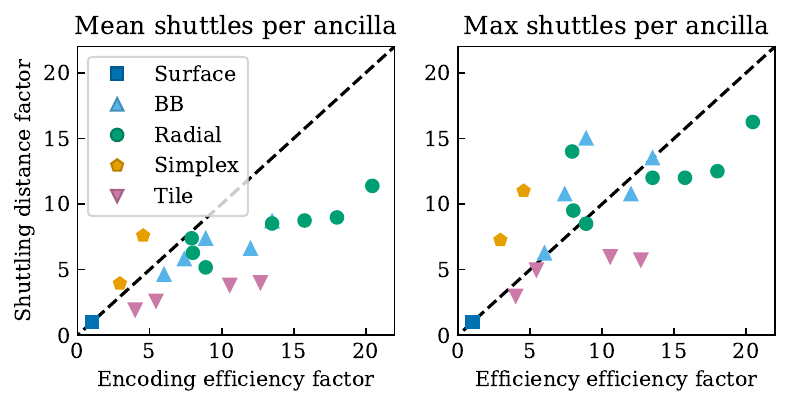}
    \caption{Comparing shuttling overhead to efficiency factor $kd^2/n$. Higher-efficiency codes from the Radial and Tile families improve encoding efficiency by more than their increase in shuttle count.}
    \label{fig:shuttle-efficiency}
\end{figure}

Scheduling these codes with \name{} is as simple as specifying the X and Z checks and the fixed positions of the data qubits. For the data qubit layout, we use the standard canonical rectangular layouts for the surface, tile, BB, and simplex codes. For radial codes and other codes without canonical layouts, we place the data qubits in a $\sqrt{n} \times \sqrt{n}$ square grid where data qubit $i$ is mapped to coordinates $\big(i \Mod{\sqrt{n}}, \lfloor i/\sqrt{n}\rfloor\big)$. It is interesting future work to consider optimizing the placement of these data qubits to further improve scheduling.

\subsection{Shuttle distance comparison}

Figure~\ref{fig:shuttle-count} shows the average number of unit-edge shuttles per ancilla qubit across the studied code instances. We compare the \name{} optimized schedules against two baselines. First, the ``ideal collision-free'' schedule assumes that shuttled qubits are allowed to occupy the same shuttling channel and even pass by one another without collisions. Compared to this baseline, the \name{} schedules (which prevent simultaneous occupation of any shuttling channel) incur a geomean 12.5\% shuttling overhead, demonstrating surprisingly near-ideal shuttling counts. For the Tile and BB code families, we also compare to ``lockstep'' schedules, in which all X and Z ancilla qubits move together synchronously \cite{viszlai_matching_2023}. These code families admit lockstep scheduling because the X and Z checks are translationally identical. \name{} outperforms these lockstep schedules by geomean 19\% and 86\%, respectively. For Tile codes, the intuitive reason for the improvement is that the X and Z checks on the edges of the array are incomplete, yet their ancilla qubits perform the same number of shuttles as the bulk checks, raising the average shuttle count. Similar edge effects lead to the large lockstep shuttle count for BB codes, but in this case the shuttle count is much higher because checks near the edge wrap around to the opposite side.

In Figure~\ref{fig:shuttle-efficiency}, we plot the relative shuttles per ancilla qubit in \name{} schedules compared to the surface code, shown against the encoding efficiency factor $kd^2/n$. This metric is a convenient way to express the overall improvement that a QLDPC code achieves over the surface code (which has $kd^2/n=1$ for all $d$), either by improving the encoding rate $k/n$ or by increasing the code distance $d$. We see in Figure~\ref{fig:shuttle-efficiency} that most of the studied codes achieve a high efficiency factor with only a modest increase in the mean relative shuttling distance, with Tile codes seeming to give the best tradeoff. Although the maximum shuttle counts are much higher, we expect the mean to be more representative of the impact on the logical error rate.

\subsection{Logical performance}

However, while $kd^2/n$ can be a useful overall metric when considering long-term scalability, the code distance $d$ does not fully capture the logical performance of a code in finite, realistic noise regimes. We therefore use Stim \cite{gidney_stim_2021} to perform circuit-level simulations of logical memory experiments for the studied code instances. For each code, we convert the compiled \name{} schedule for $d$ rounds of syndrome extraction into a Stim circuit, adding error instructions after each initialization, gate, shuttle sequence, or readout. We also add idling errors on qubits during inactive periods, such as while a data qubit waits between ancilla visits. The operations and associated error rates are summarized in Table~\ref{tab:ops}. We use constant $T_1 =$ \SI{10}{\second} and $T_2 =$ \SI{10}{\milli\second} (assuming dynamical decoupling techniques are used during any idle periods).

\begin{table}[]
    \centering
    \begin{tabular}{llll}
        Operation & Duration & Error rate & Noise structure \\
        \hline
        CX gate & \SI{100}{\nano\second} & $10^{-3}$ & Depolarizing \\
        H gate & \SI{100}{\nano\second} & $10^{-3}$ & Depolarizing \\
        Initialization & \SI{500}{\nano\second} & $10^{-3}$ & Bit-flip ($X$) \\
        Measurement & \SI{500}{\nano\second} & $10^{-3}$ & Bit-flip ($X$) \\
        Shuttle (one edge) & \SI{1000}{\nano\second} & $10^{-4} - 10^{-2}$ & Phase-flip ($Z$) \\
        Displace & \SI{200}{\nano\second} & $10^{-4} - 10^{-2}$ & Phase-flip ($Z$) \\
        Idle decoherence & $t$ & $1-e^{-t/T_1}$ & Bit-flip ($X$) \\
        Idle dephasing & $t$ & $1-e^{-t/T_2}$ & Phase-flip ($Z$) \\
    \end{tabular}

    \hspace{0.5ex}
    
    \caption{Operations and associated error rates for circuit-level simulation}
    \label{tab:ops}
\end{table}

\begin{figure}
    \centering
    \includegraphics[width=\linewidth]{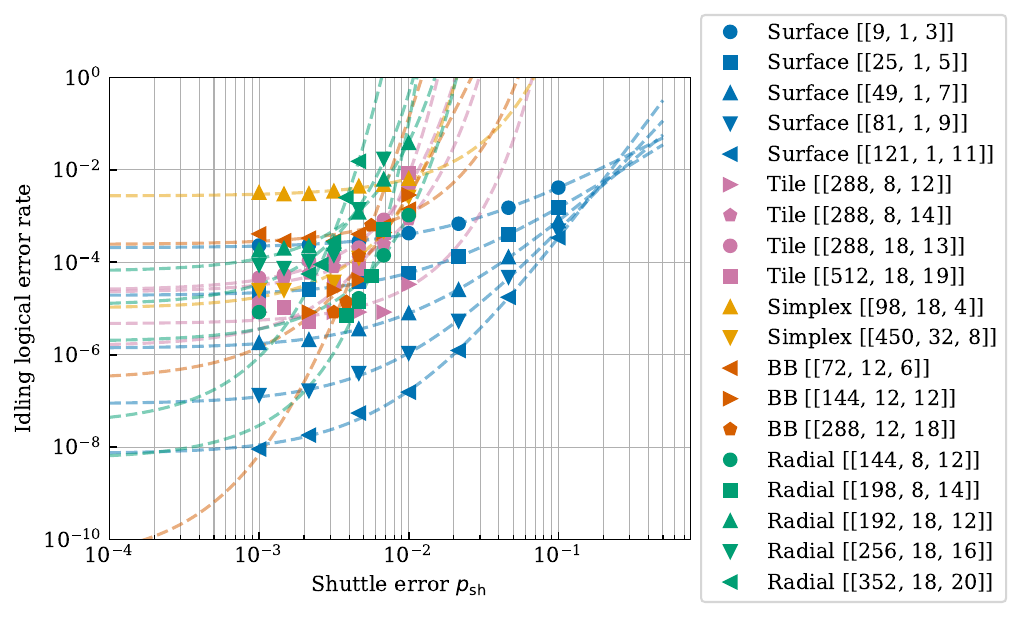}
    \caption{Simulated logical error rate data and corresponding model fits for the code instances studied in this work, as a function of the shuttling error rate $p_\text{sh}$. All other error rates are kept constant (see Table~\ref{tab:ops}). Reducing the shuttling error rate below $p_\text{sh} \approx 10^{-3}$ yields dimishing returns, as the codes are generally limited by the constant gate error rates.}
    \label{fig:sim-fits}
\end{figure}

\begin{figure*}
    \centering
    \includegraphics[width=0.9\linewidth]{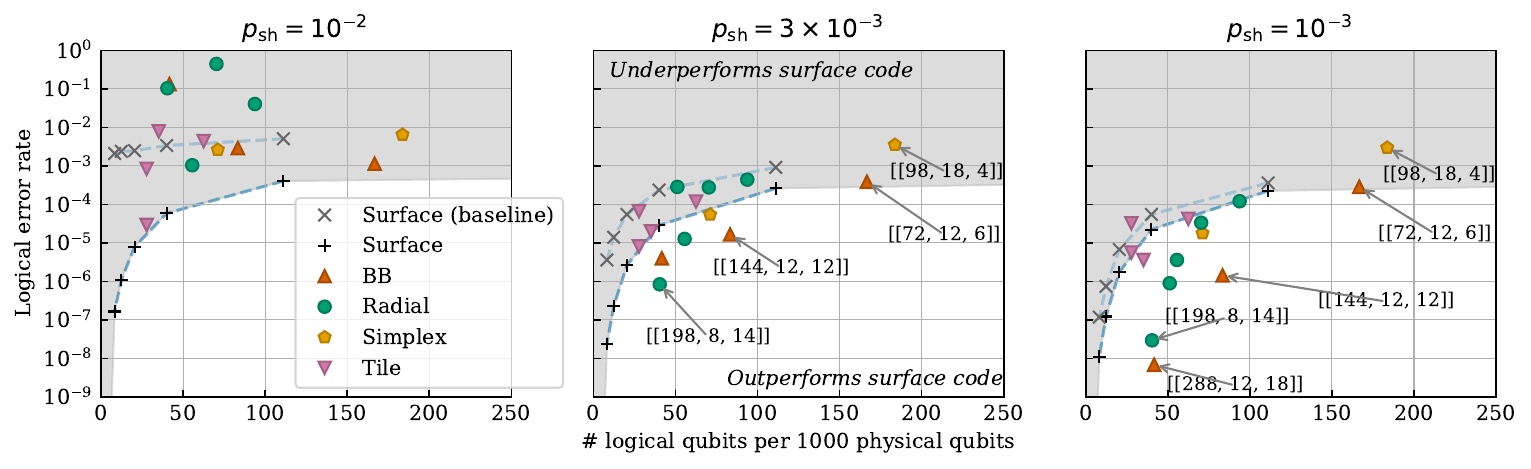}
    \caption{Comparing \name{}-optimized QLDPC and surface code performance. X axis shows the number of logical qubits that can be encoded in 1000 physical qubits (equivalent to $1000k/n$), and Y axis shows the idling logical error rate per logical qubit. Each plot shows a different value of $p_\text{sh}$; all other physical error rates are held constant. Surface code points are connected by a dashed line, and the unshaded region below the line indicates the regime where QLDPC codes outperform the \name{}-optimized surface code in either encoding efficiency or logical protection. Also shown is the baseline surface code, which does not include circuit tailoring (Section \ref{sec:tailoring}). For $p_\text{sh} = 10^{-2}$ (top), no studied QLDPC code outperforms the \name{} surface code, but at $p_\text{sh}=3 \times 10^{-3}$ (center) we identify several codes that provide up to 10$\times$ reduced logical error rates compared to the surface codes with similar encoding rates. For $p_\text{sh} = 10^{-3}$ (bottom), the improvements are much stronger, with the $[[288,12,18]]$ BB code providing approximately 3200$\times$ improvement in logical error rate compared to the surface code with similar encoding rate, and a 5$\times$ improvement in encoding rate compared to the surface code with similar logical error rate.}
    \label{fig:encoding-rate-vs-ler}
\end{figure*}

We perform simulations with all error rates constant except the shuttling and displacement error rates, which we vary between $10^{-3}$ and $10^{-1}$. We use the noise-aware circuit tailoring technique described in Section~\ref{sec:tailoring} for all schedules, including the surface code. For some codes, the logical error rates were too low to measure directly with Monte Carlo simulation methods, so we perform fits to the data for higher $p_\text{sh}$ and extrapolate to lower $p_\text{sh}$ as needed. We fit the theoretical model $p_L = A(b+c\cdot p_\text{sh})^{d_{\text{eff}}}$, where $A,b,c,$ and $d_{\text{eff}}$ are fit parameters. Figure~\ref{fig:sim-fits} shows the simulation data and the obtained fits. For most codes, we see diminishing returns beyond $p_\text{sh} \approx 10^{-3}$, as gate errors begin to limit performance. While we are unable to directly observe lower logical error rates, the asymptotic behavior of the fits (in the limit of negligible shuttling error) roughly aligns with expected logical error rates under the constant $p=10^{-3}$ gate and measurement errors \cite{bravyi_highthreshold_2024, scruby_highthreshold_2024, malcolm_computing_2025, steffan_tile_2025}.

In Figure~\ref{fig:encoding-rate-vs-ler}, we plot the logical error rates (extrapolated where necessary) at specific shuttling error rates $10^{-2}, 3 \times 10^{-3},$ and $10^{-3}$. The $x$ axis shows the number of logical qubits that are encoded per 1000 physical qubits (calculated as $1000 k/n$). We connect the surface code points with a dashed line to form a frontier and shade the area above. The unshaded region therefore corresponds to the area of net improvement over surface codes. A QLDPC data point in the unshaded region can therefore be viewed in one of two ways. Comparing to the surface code frontier above it, the QLDPC code can be seen as improving the logical error rate at an equivalent error rate. Alternatively, comparing to the surface code frontier to the left, the QLDPC code can be seen as improving the encoding rate at the same logical error rate. For $p_\text{sh} = 10^{-2}$, we see that no QLDPC codes are able to improve upon the surface code. However, at $p_\text{sh} = 3 \times 10^{-3}$, several BB and Radial codes enter the successful regime, and further improvements are obtained at $p_\text{sh}=10^{-3}$. Taking the example of the $[[144,12,12]]$ BB code at $p_\text{sh}=10^{-3}$, we can see that at the same logical error rate it improves upon the surface code encoding rate by approximately $4\times$, from $20$ to $83$; alternatively, at the same encoding rate, it reduces the logical error rate per logical qubit by two orders of magnitude, from $10^{-4}$ to $10^{-6}$. We emphasize that these results are only possible with the circuit tailoring technique. Without circuit tailoring, no QLDPC codes outperform the surface code even at $p_\text{sh}=10^{-3}$ due to the accumulation of propagated errors to the data qubits. 

\begin{figure}
    \centering
    \includegraphics[width=0.9\linewidth]{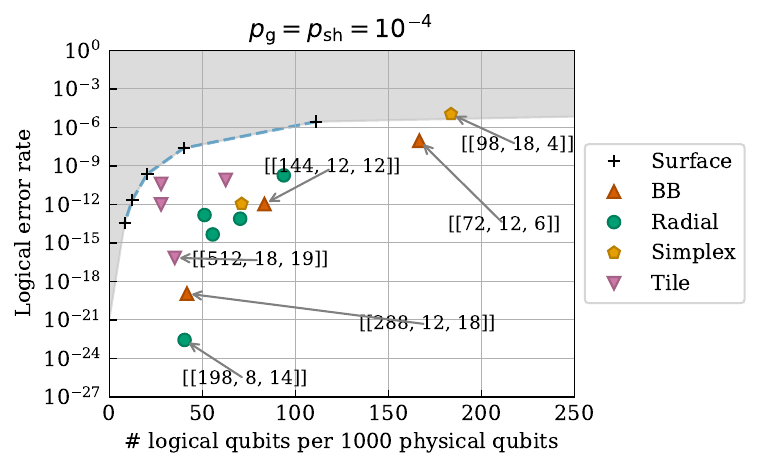}
    \caption{Lowering physical gate and shuttling error rates to $10^{-4}$ enables strong improvements in performance, dramatically lowering logical error rates.}
    \label{fig:lower-gate-err}
\end{figure}

Improving $p_\text{sh}$ below $10^{-3}$ does not significantly improve the performance of most codes because the constant gate error rates become the limiting factor. In Figure~\ref{fig:lower-gate-err}, we investigate the case where $p_{CX}$ and $p_H$ are reduced to $10^{-4}$. This enables very strong performance at lower shuttling error rates. For $p_\text{sh}=10^{-4}$, we see strong gains for many of the studied codes. The $[[288,12,18]]$ BB code achieves approximately 8 orders of magnitude reduction in logical error rate compared to the surface code with the same encoding rate. However, reaching such low gate and shuttle error rates is hypothetical; we show these results for completeness, but in practice these low noise levels may not be reachable on a scaled-up physical device.

\section{Discussion}

In this work, we introduced \name{}, a comprehensive framework for efficiently scheduling and implementing high-rate QLDPC codes on shuttling-based semiconductor spin qubit architectures. By addressing both the physical noise mechanisms of shuttling and the algorithmic challenges of ancilla routing, we bridge the gap between near-term hardware capabilities and the non-local connectivity requirements of advanced quantum error correction.

First, we demonstrated that a simple, noise-aware circuit tailoring technique---appending Hadamard gates around shuttling operations for X ancilla qubits---effectively mitigates the dominant dephasing noise inherent to spin qubit shuttling. This prevents error propagation to the data qubits, relaxing the viable shuttling error threshold by 5-10$\times$ and extending the feasible range of ancilla movement.

Building upon this increased resilience, we developed Q-SIPP, a novel path-planning algorithm adapted from classical robotics to navigate the dynamic obstacle space of mobile ancilla qubits. Q-SIPP achieves near-ideal scheduling with only a 12.5\% overhead compared to collision-free ideals. Furthermore, it significantly outperforms manually designed lockstep schedules, reducing shuttling overhead by up to 86\% for code families with complex edge boundaries such as bivariate bicycle codes.

Through detailed circuit-level simulations, we identified the crossover regime in which these optimizations allow QLDPC codes to definitively surpass the surface code on this architecture. At shuttling error rates of $p_\text{sh} \leq 3\times 10^{-3}$, families such as the bivariate bicycle and radial codes begin to offer drastic improvements. Notably, at $p_\text{sh}=10^{-3}$, the [[144,12,12]] bivariate bicycle code achieves a 4$\times$ higher encoding rate or a 100$\times$ reduction in logical error rate compared to an equivalent surface code.

There are several promising avenues for future work. As noted in Section~\ref{sec:eval}, our evaluation utilized simple grid layouts for data qubits; explicitly co-optimizing the spatial mapping of data qubits with the \name{} routing engine could further minimize shuttling bottlenecks. Additionally, while the exact TSP solver within Q-SIPP's heuristic is highly efficient for the check weights evaluated in this work (up to weight 8), integrating approximate TSP solvers or designing alternative Q-SIPP heuristic functions will be necessary to schedule codes with significantly higher check weights. Finally, applying a modified Q-SIPP algorithm to heterogeneous chip layouts, such as architectures with dense data clusters connected by sparse long-range shuttling channels, remains an exciting direction.

Ultimately, our results indicate that the lack of native long-range interactions in spin qubits is not a fundamental roadblock for high-efficiency quantum error correction. With coordinated, collision-aware scheduling and tailored circuits, shuttling-based architectures present a highly scalable and viable path toward large-scale fault-tolerant quantum computation.

\bibliographystyle{unsrt}
\bibliography{references}

\end{document}